\documentclass{style}
\usepackage{graphicx}

\newcounter{bla}

\begin{document}
\begin{frontmatter}

\title{{\textit{JuNoLo}} - J\"ulich Non Local code for parallel calculation of vdW-DF nonlocal density functional theory  }

\author[a]{Predrag Lazi\'{c}\thanksref{author}\thanksref{leave}},
\author[a,d]{Nicolae Atodiresei},
\author[a]{Mojtaba Alaei},
\author[b]{Vasile Caciuc},
\author[a]{Stefan Bl\"ugel},
\author[c]{Radovan Brako}

\thanks[author]{Corresponding author}
\thanks[leave]{On leave of absence from the Rudjer Bo\v{s}kovi\'{c} Institute, Zagreb, Croatia.}

\address[a]{Institut f\"ur Festk\"orperforschung (IFF) and Institute for Advanced Simulation (IAS), Forschungszentrum J\"ulich, 52425 J\"ulich, Germany}
\address[b]{Physikalisches Institut, Westf\"alische Wilhelms Universit\"at M\"unster, Wilhelm-Klemm-Str.~10, 48149 M\"unster, Germany}
\address[c]{Rudjer Bo\v skovi\'c Institute, P.O. Box 180, 10002 Zagreb, Croatia} 
\address[d]{The Institute of Scientific and Industrial Research, Osaka University, 567-0047 Osaka, Japan}

\begin{abstract}
  %Type your abstract here.
Nowadays the state of the art Density Functional Theory (DFT) codes are based on local (LDA) or semilocal (GGA) energy functionals. Recently the theory of a truly nonlocal energy functional has been developed. It has been used mostly as a post DFT calculation approach, i.e. by applying the functional on the charge density calculated using any standard DFT code, thus obtaining a new improved value for the total energy of the system. Nonlocal calculation is computationally quite expensive and scales as $N^2$ where $N$ is the number of points in which charge density is defined, and a massively parallel calculation is essential for a wider applicability of the new approach. In this article we present a code which acomplishes this goal.

\begin{flushleft}
  %Insert your suggested PACS number here
\end{flushleft}

\begin{keyword}
Electronic structure; Density functional theory; Van der Waals interaction; nonlocal correlation. 
  % Please give some freely chosen keywords that we can use in a
  % cumulative keyword index.
\end{keyword}

\end{abstract}

\end{frontmatter}

\newpage

% In program descriptions the main text of the paper is listed under
% the heading LONG WRITE-UP.

\hspace{1pc}
%{\bf LONG WRITE-UP}

\section{Introduction}
In recent years the codes based on the density functional theory (DFT) \cite{DFT} have been the main tool for exploring theoretically the properties of materials. 
Despite the large success of the approach, there is a class of systems for which the present codes fail miserably. The main reason is that the energy functionals used are based on the local (LDA) or the semilocal approximations (GGA) which neglect the long range correlation contribution to the total energy. The systems in which DFT codes fail are mostly those in which the van der Waals energy makes a significant contribution to the total energy of the system. Recently Dion et al. \cite{dion-prl} proposed the first fully nonlocal energy functional (vdW-DF) based on first principles which could be easilly implemented. Originally, this seamless theory was tested as a DFT postprocessing tool that yields the new value for total energy. Justification for such approach came later \cite{thonhauser} when the theory was implemented selfconsistently into a DFT code, allowing Kohn-Sham eigenfunctions to change. The result of the self consistent calculations made so far confirm that it suffices to use vdW-DF functional as a post DFT perturbation because the changes in K-S eigenfunctions are negligible. Consequently, one should accept the vdW-DF (even in the postprocessing implementation) to be on an equal footing with the well established LDA and GGA functionals. 
The value of the first nonlocal functional has been well recognized judging on the number of citations of the original paper \cite{dion-prl}, but until now only relatively small systems have been treated with it. We guess that the main reasons for this are the high numerical cost of the calculations and nonexistence of a reliable and widely available parallel code. The main calculation within the seamless theory can be nicely parallelized and with the parallel code that we present here one can treat any system that is solvable with present DFT codes. Moreover, our code is applicable no matter which DFT code was used to obtain charge density. One can apply the code to the results of DFT calculations done in a plane wave basis, some other basis, in real space implementation, LCAO, or any other technique. This makes the code interesting to a large community.\\

A historical timeline of the {\textit {JuNoLo}} code is rather brief:\\
\\
$\bullet$ {\bf 2005-2007:} Two of the authors (R.~B. and P.~L. at Rudjer Boskovic Institute in Zagreb, Croatia) started with implementation of the vdW-DF theory in Python. The code was serial but the calculation of the kernel was also implemented in parallel code in Fortran90. 
The code was succesfully tested on several examples, in particular on the Kr dimer (Fig. 1 in Ref. \cite{JPC} ). \\
$\bullet$ {\bf 2007-2008:} Using the past experience the code was writen from scratch in Fortan90 at Forschungszentrum J\"ulich in Germany, in a form suitable
 for massively parallel calculations.

The remainder of the paper is organized as follows. In the next section we give a basic theoretical background of the nonlocal correlation energy, i.e. of the vdW-DF functional. In Section 3 we describe the structure of the {\textit {JuNoLo}} program following the flow chart.  In Section 4 we give brief installation instructions, followed by explanations on running the program in Section 5. In Section 5 we also address the speedup properties of our program.  In Section 6 we provide the results for two test calculations, the Krypton dimer and the Xenon monolayer.  Finally, in Section 7 we give conclusions. In the Appendix the full listings of a sample input and output files are given.

\section{Theoretical background}
For the fine details of the seamless theory, i.e. vdW-DF functional, one should consult Ref. \cite{dion-prl} and references therein. Here we give only a brief overview necessary to explain our code. The name {\textit seamless} for the theory reflects its behavior in extreme regimes of application.  Suppose that we have two well separated fragments of the system - the molecule (adsorbate) and the surface (substrate). At very large distances between the two only pure van der Waals forces are at work, which can be described accurately also by a semiempirical theory. The seamless theory describes this situation correctly as well. As the molecule approaches the surface the chemical bonding starts to take place, and when the molecule eventually bonds strongly to the surface the semilocal (GGA) energy functional describes the situation correctlyi (in vast majority of cases, at least). The application of seamless theory to this case will not spoil the results. But the seamless theory is tailored to be valid also in between these two extreme situations, hence the name. Moreover, recently it has been realized that the vdW-DF contribution can play a major role even in a covalently bonded system \cite{Johnston}, making the area for the application of the {\textit {JuNoLo}} code much larger.

The key ingredient for application of the seamless theory is the charge density $n({\bf r})$ that one obtains from a DFT calculation.
The charge density must be given on a real space grid, with equidistant division in all three spatial dimensions. The axes need not to be orthogonal. The total nonlocal correlation energy is given by the integral:
\begin{equation}
\label{integral}
E_{\mathrm{c}}^{\mathrm{NL}}=\frac{1}{2}\int\int d^3{\bf r}d^3{\bf r'}n({\bf r})\phi({\bf r},{\bf r})n({\bf r'}).
\end{equation}
%In calculation of integral (\ref{integral}) $n({\bf r})$ is a charge density at point ${\bf r}$ and $\phi({\bf r},{\bf r'})$ is the kernel. 
The kernel $\phi({\bf r},{\bf r'})$ in (\ref{integral})  can be redefined to depend on two quantities, namely $D\geq 0$ and $0\leq \delta < 1$ as described in Ref. \cite{dion-prl}. The kernel $\phi$ is calculated according to formula (14) in Ref. \cite{dion-prl}, which requires a double integration.
For large values of $D$ an asymptotic expression for $\phi$ exists, given by formula (17) in Ref. \cite{dion-prl}.
The bulk of our program is concerned with the calculation of the double integral (\ref{integral}) which in numerical calculations reduces to a discrete double sum:
\begin{equation}
\label{suma}
E_{\mathrm{c}}^{\mathrm{NL}}=\frac{1}{2} \sum_{\rm i} \sum_{\rm j} n(r_{\rm i}) \phi(r_{\rm i},r_{\rm j}) n(r_{\rm j}).
\end{equation}
From the values of the total energy obtained in a DFT calculation $E_{\mathrm{tot}}^{\mathrm{DFT}}$ and the calculated $E_{\mathrm{c}}^{\mathrm{NL}}$ one gets the new value of the total energy:
\begin{equation}
\label{nonlocal}
E_{\mathrm{tot}}^{\mathrm{NL}}=E_{\mathrm{tot}}^{\mathrm{DFT}}-E_{\rm x}^{\mathrm{PBE}}-E_{\rm c}^{\mathrm{PBE}}+(E_{\rm x}^{\mathrm{PBE}}+E_{\rm c}^{\mathrm{LDA}}+E_{\rm c}^{\mathrm{NL}}).
\end{equation}
Here the subscripts x and c denote various contributions to the exchange and correlation energy, respectively. Thus $E_{\rm x}^{\mathrm{PBE}}$ and $E_{\rm c}^{\mathrm{PBE}}$ are the GGA exchange and correlation contributions (both local and semilocal terms), assuming that the PBE flavor of hte functional has been used, and $E_{\rm c}^{\mathrm{LDA}}$ is the LDA correlation (i.e. the local term only). In this picture the correlation contribution has been subtracted from the total DFT energy so that the seamless nonlocal term can be added. The {\textit {JuNoLo}} code calculates all the aforementioned energy contributions from the provided charge density, which is particularly useful if the user's DFT code does not write those contributions separately. In addition the {\textit {JuNoLo}} code also yields the value of the exchange energy according to revPBE energy functional \cite{revPBE}. It has been claimed in literature that $E_{\rm x}^{\mathrm{revPBE}}$ may be a better choice in the vdW-DF calculations, so this value may be used as the fourth term in (\ref{nonlocal}) (which is the reason why $E_{\rm x}^{\mathrm{revPBE}}$ has been subtracted in the first place). If another functional is used in DFT calculation, such as PW91 \cite{PW91} user should take care how to obtain needed values for formula (\ref{nonlocal}). 

Also a word of advice is in order here. In the case that a real physisorbed system is calculated, i.e. only van der Waals bonding is important with little or no chemical bonding, one should consider using a different implementation of vdW-DF theory. In such cases one has more or less well separated parts of the system which enables localization of charge using tools such as Wannier functions and obtaining integral values at much lower costs, see for example Ref. \cite{Silvestrelli}. In such cases of fragmented systems one could consider application of much simpler theories alltogether, such as semiempirical van der Waals implementation \cite{semiempirical}. The {\textit {JuNoLo}} code is intended for the most general usage of vdW-DF theory, especially in cases where chemical bonding is taking place so that one can not tell apart fragments of the bonded system. It has been recently shown that vdW-DF theory can significantly influences the results for chemisorbed systems \cite{Johnston,Lazic}.

\section{Description of the code}
\begin{figure}[htb]
\begin{center}
\includegraphics[scale=0.5,clip=true]{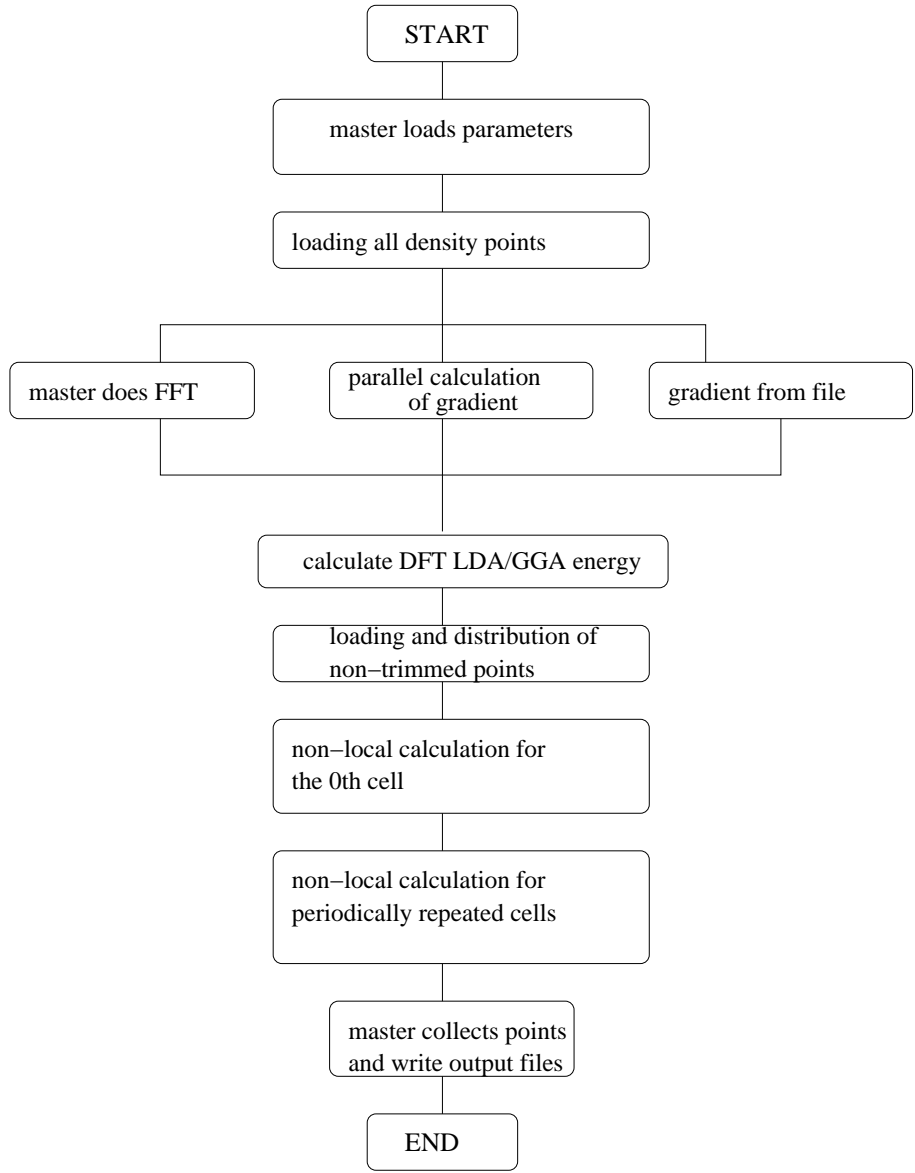}
\end{center}
\caption{Flowchart of the {\textit {JuNoLo}} program.}
\label{flow}
\end{figure}

The flow chart of the {\textit {JuNoLo}} program is given in Fig. \ref{flow}.
For the fully detailed description of the program's internal structure and functioning the reader is referred to the documentation provided in the {\textit {JuNoLo}} distribution \cite{JuNoLo}. 
A brief description of steps from the flow chart is given here. \\ 
$\bullet$ The master CPU loads parameters from the \tt input \rm file and distributes them to all CPUs participating in the calculation.
Whenever such communication is needed in which all CPUs have to get data from the master, communication through the binary tree structure is used.\\
$\bullet$ The master loads the density from the charge density file provided. Based on the choice made in the \tt input \rm file the calculation of the gradient at all points is done either by means of FFT (using a FFTW library, done only on the master processor), real space finite differences approach implemented for parallel computing (perpendicular axes only), or the gradient values are loaded from a text file (this file is generated every time the calculation is done with the FFT option selected).\\
$\bullet$ Calculation of the LDA and GGA energy is performed, even though this is not necessary if user can obtain this energy values from the original DFT code. However this is not always the case so therefore we calculate also these contributions.\\
$\bullet$ Due to a small significance of low density points to the vdW-DF energy we introduce a cutoff density value below which the density is not considered in the calculation. This does not reduce the accuracy of the energy value but can reduce the time needed for numerical calculations tremendously, in particular for open systems with surfaces. \\
$\bullet$ The points containing density larger than the cutoff value are distributed among CPUs in a balanced manner, i.e. in ideal situation each CPU would be assigned the same number of points. In our implementation the largest difference in number of points between CPUs is one. \\
$\bullet$ The calculation of the nonlocal correlation, i.e. vdW-DF functional value, is performed within the single unit cell (0th cell - the one described in charge density file). During this process CPUs send their points to other CPUs (and receive points from other CPUs of course).\\
$\bullet$ If the problem is of periodical nature, like Xenon monolayer example given in a later section, the charge density from the 0th cell must be repeated in space and the interaction with this repeated charge density is calculated. For details the user should consult the manual.\\
$\bullet$ The master processor collects information from all CPUs and writes final output files.\\

\section{Installation instructions}
{\textit {JuNoLo}} is distributed as a gzipped tar file. The program manual and few examples are available on the web page \cite{{JuNoLo}}. On Linux platforms it should be unpacked by typing\\
\tt > tar -xvzf junolo.tar.gz \rm \\
which will create a directory containing the program source code.
In the program source directory \tt VDW\_SRC \rm several makefiles are provided with the names \tt Makefile.platform \rm.
The values of system dependent parameters defined in the Makefile such as the name of the fortran90 compiler, optimization flags and location 
of the fftw library should be changed by user according to his particular system.
The compilation of the executable program is done by a command\\
\tt > make -f Makefile.platform \rm\\
which results in a \tt vdw.exe \rm executable file.
%Some further details can be found in the \tt README \rm file in the \tt VDW\_SRC \rm directory.

\section{Running the program}
Once compiled {\textit {JuNoLo}} can be run as a single processor program or as a multiprocessor program. 
The prerequisites to run the code is that the user has already done a DFT calculation and obtained the charge density from it in a format 
appropriate for the {\textit {JuNoLo}} code. In the examples package we have provided three Python scripts that prepare the needed charge density file from 
the files generated by standard DFT codes: DACAPO \cite{dacapo,dacapo2}, PWscf \cite{pwscf} and VASP \cite{vasp,VASP1,VASP2}. In the program manual it is described 
how one can prepare the charge density file on his own.
The user also needs to provide the \tt input \rm file in which calculation parameters are specified. For a detailed explanation of each parameter the user should 
consult the program manual. The \tt input \rm file and the charge density file should be placed in the same directory. Usually the kernel file is also needed in the same directory (except if the option in the \tt input \rm file is set to calculate the kernel).\\
Once having all the mentioned files in the same directory the user should give the command:\\
\\
\tt \noindent ./vdw.exe input \rm \\
\\
\noindent for a single processor run, or\\ 
\\
\tt \noindent mpirun -np 16 ./vdw.exe input \rm \\
\\
\noindent for a run on 16 CPUs, for example.

We have performed the timing of the code on an Intel core Duo 2.13 GHz processor with 2 Gb of RAM and Linux OS. 
The calculation containing 130,000 points (after trimming) takes 36 minutes on a single CPU core. The running time scales as $N^2$ with number of points.

\subsection{Speedup}
\begin{figure}[htb]
\begin{center}
\includegraphics[scale=0.5,clip=true]{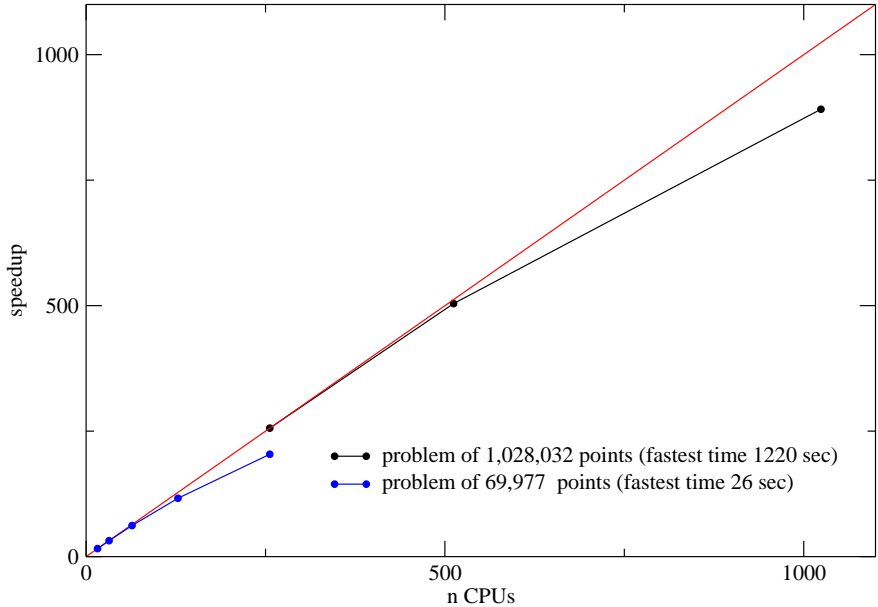}
\end{center}
\caption{Speedup curves for the {\textit {JuNoLo}} program. Red line represents the ideal speedup, linear in the number of CPUs used. Shortest time is the duration of the calculation when the largest number of CPUs is used, i.e. at the last point of the corresponding curve.}
\label{speed1}
\end{figure}

The strongest point of the {\textit {JuNoLo}} code is its scalability. We have done speedup measurements for the code on several problems of different size. 
The speedup curves for two problems are shown in Fig. \ref{speed1}.
Parallelization of the code is done by distributing over processors the points containing the charge density. The more processors we have the less points per processor 
will be assigned. However, a critical number of points per processor exists after which there is no sense in adding more processors to calculation due to a high communication/calculation time ratio per processor. For our BlueGene/P (850 MHz CPUs) system the critical number of points was determined to be 1500. Before reaching the critical number the speedup is practically linear. A very detailed description of parallelization in the code is given in the manual.
Besides the well behaved speedup, memory consumption per CPU drops exactly linearly with the number of CPUs used.
The test run of the code on the full BlueGene/P system of 65536 CPUs was successfully performed.

\section{Two examples}
In this section we provide two example calculations. Both examples represent weakly bonded systems, i.e. systems bonded through van der Waals interactions rather than chemical bonds, and due to that are not suitable for treatment with present DFT codes based on local (LDA) or semilocal (GGA) functionals. The reason that we present both is due to a clear difference between them, namely Krypton dimer is a nonperiodical system while Xenon layer is a periodical (infinite) system. The {\textit {JuNoLo}} code is capable of handling both types of problems.
\subsection{Krypton dimer}
\begin{figure}[htb]
\begin{center}
\includegraphics[scale=0.3,clip=true]{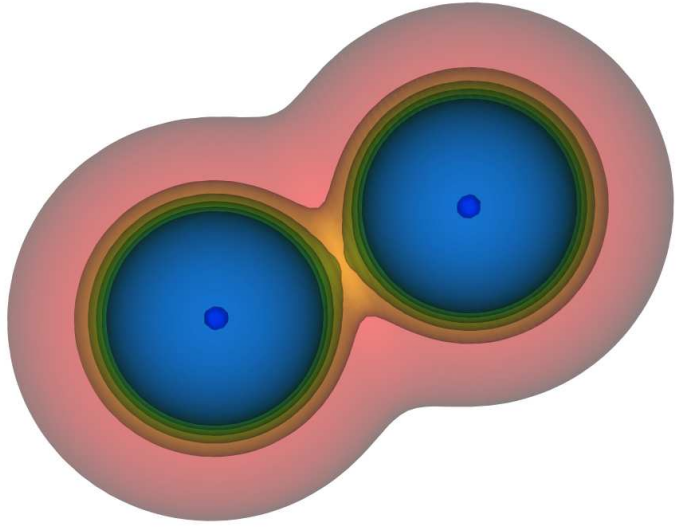}
\end{center}
\caption{Isosurfaces of a Krypton dimer charge density. The outermost isosurface corresponds to a very low charge density.}
\label{krdimercharge}
\end{figure}

Krypton dimer is, due to the fact that Krypton is a noble gas, very weakly chemically bonded. The bonding between Krypton atoms is originating mostly from the van der Waals interaction i.e. from nonlocal correlation. Because of this fact the calculation of Krypton dimer by present DFT codes which use local or semilocal functionals gives unphysical result. That makes this system a perfect candidate for a test of the {\textit {JuNoLo}} code. Also, this system has already been calculated by the authors of the vdW-DF theory \cite{dion-prl} and therefore can be used as a benchmark for the correctness of our code.
Our DFT calculations for this example were done with DACAPO and VASP code. Both are plane wave codes with a major distinction that VASP uses PAW theory \cite{PAW} while DACAPO does not. Calculation in VASP was done using the PBE functional selfconsistently, while DACAPO uses PW91 functional but we recalculated the PBE energy values from the resulting densities. We also used selfconsistent charge densities obtained by both codes to recalculate revPBE energies (nonselfconsistently, of course, but this procedure can be justified).
In DACAPO there is a possibility to use the valence charge only or to add the core charge. We did both calculations and the results for the energy differences were the same. 
%However this is not the general rule. 
In VASP we used only valence charge as input for our vdW-DF calculation.  
The case of a Krypton dimer is intrinsically nonperiodic and therefore less numerically expensive for calculation than Xenon monolayer. The charge density of Krypton dimer is given in Fig. \ref{krdimercharge}.

\begin{figure}[htb]
\begin{center}
\includegraphics[scale=0.5,clip=true]{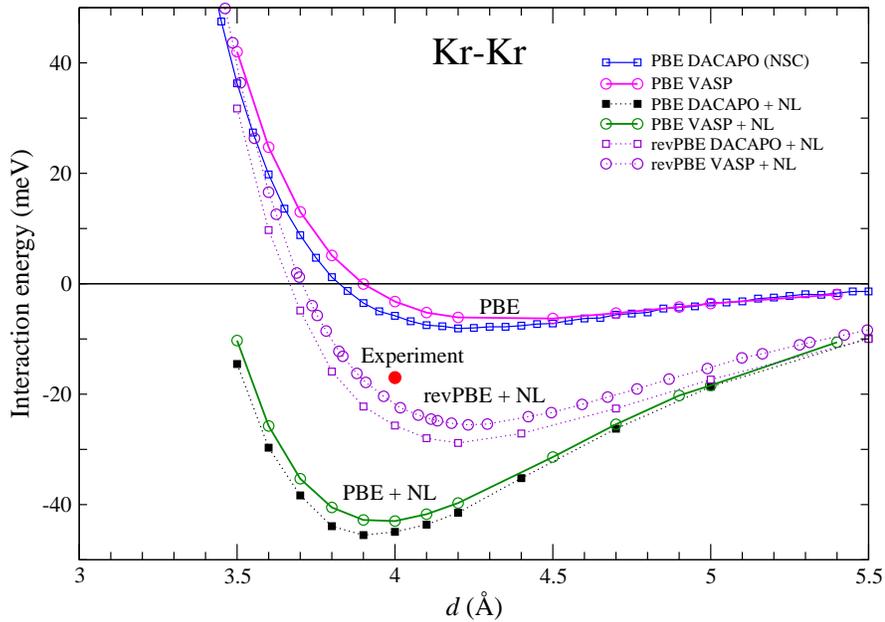}
\end{center}
\caption{Kr dimer binding energy in DFT GGA and vdW-DF calculation. PBE, revPBE+NL and PBE+NL labels are positioned in the vicinity of the corresponding minima obtained for this energy functional. NL means that nonlocal correlation energy was added, i.e. PBE+NL means that it is a PBE DFT calculation on top of which vdW-DF theory has been applied. Agreement with previous results of Dion et al. \cite{dion-prl} is excellent.}
\label{krdimerenergy}
\end{figure}

Calculating the energy values for several different distances between atoms in a dimer we obtained the binding energy curve shown in Fig. \ref{krdimerenergy}.

As can be seen from the Fig. \ref{krdimerenergy} pure DFT results are far away from the experimental values for the interatomic distance and interaction energy. Results for PBE+NL give a correct equilibrium interatomic distance but the binding is too strong. In revPBE+NL case the binding energy seems better but the interatomic distance is somewhat too large. As the vdW-DF theory addresses only correlation effects the question of the exchange energy remains. It has been strongly argued by the authors of the vdW-DF theory and others \cite{dion-prl},\cite{Johnston} that revPBE should be used to describe exchange energy due to spurious binding that occurs in PBE functional. However, we believe that the best way to deal with the exchange part of the energy may be to use exact exchange calculation, either to get the correct number or at least to check which of the available exchange functionals is closest to the behavior of the exact exchange for a given problem.

\subsection{Xenon monolayer}
\begin{figure}[htb]
\begin{center}
\mbox{\includegraphics[scale=0.19,clip=true]{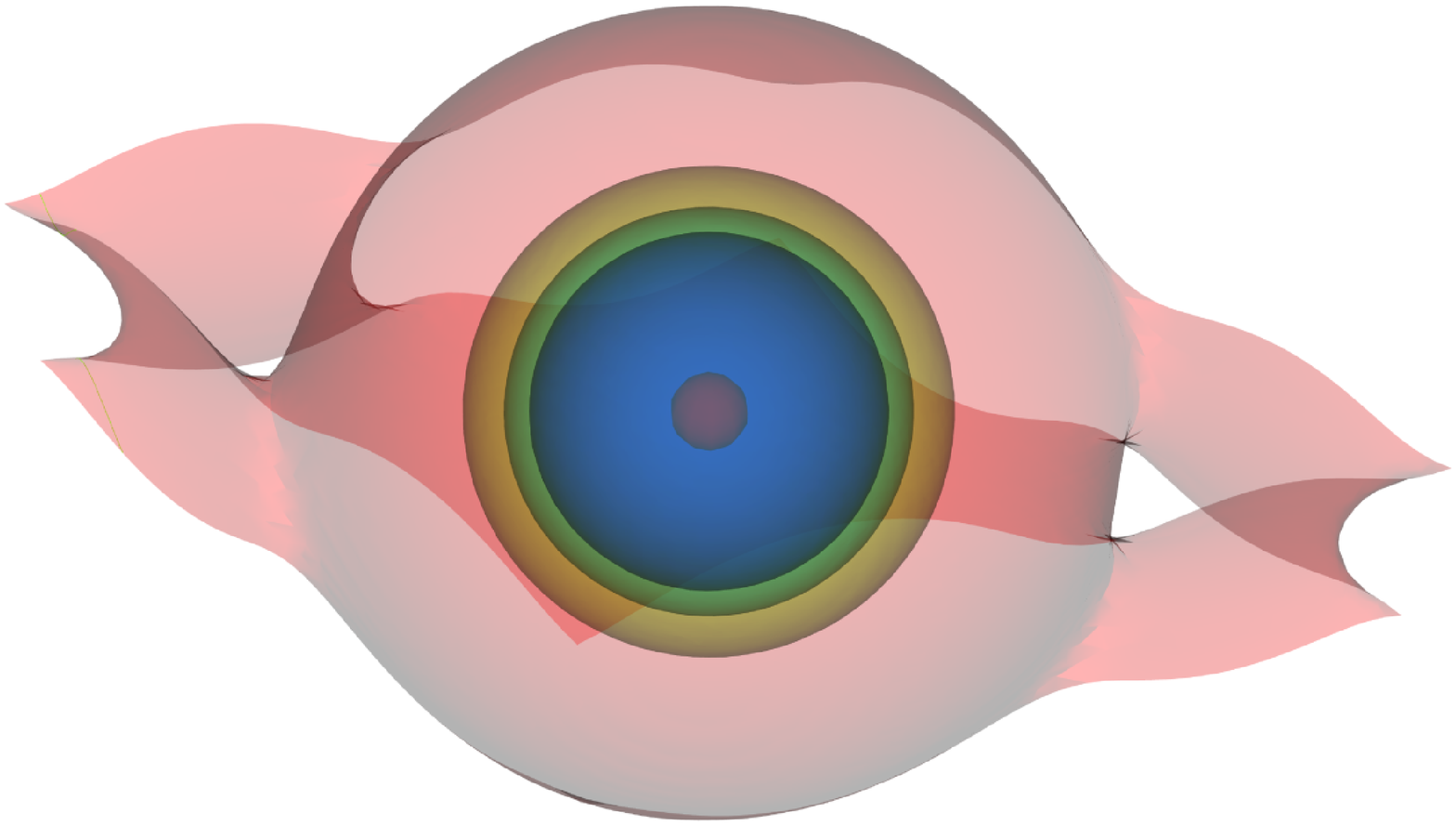}}
\mbox{\includegraphics[scale=0.19,clip=true]{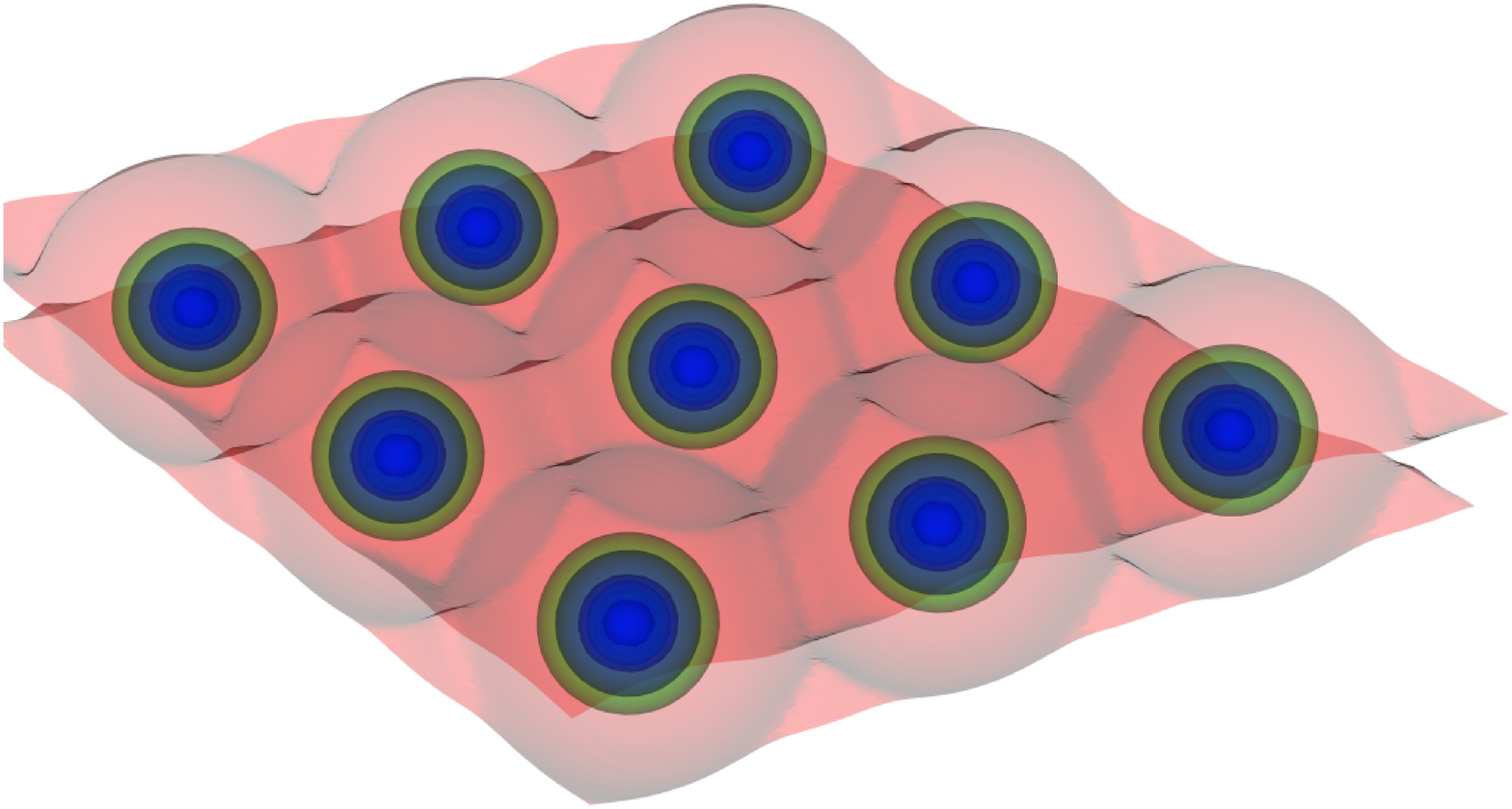}}
\end{center}
\caption{Isosurfaces of charge density used for Xenon monolayer calculation. On the left is a single unit cell, with the Xe atom situated in the middle of the cell. On the right is the density repeated several times. The outermost isosurface corresponds to a very small charge density.}
\label{xelayer}
\end{figure}

The case of self-standing Xenon monolayer in vacuum might seem a bit unrealistic because such structure does not exist and is probably impossible to build. However, growing a Xenon monolayer on Cu(111) surface can be done in such a way that Xenon makes a monolayer commensurate with the copper surface with a coverage $\theta=1/3$. Due to very weak chemical bonding between Xenon and copper surface one can deduce from the experiments the properties of a self-standing Xenon monolayer in this configuration.
The case of a Xenon monolayer is very different from the Kr dimer case due to the fact that the monolayer is in principle infinite. We have used a minimal unit cell to make this calculation using the DACAPO code. The charge density of a single unit cell and unit cell repeated several times in the monolayer plane is shown in Fig. \ref{xelayer}. 

Calculating the energy for different values of the layer lattice constants we obtain the energy minimum at around 4.4 $\mathrm{\AA}$ in very good 
agreement with experiment. From this calculation one can also analyze the curvature of the energy around the minimum and obtain vibration frequencies which are comparable to the experimental ones. This is a nontrivial result and shows that LDA calculations, which sometimes by accident yield good position of the energy minimum, should be avoided in treating this kind of systems. The reason that LDA calculations sometimes give good position of the energy minimum (or at least better than GGA functionals) in a predominantly van der Waals bonded systems is due to a simple approximations for correlation built in LDA functional which leads to overbinding, sometimes accidentally yielding a correct position of the energy minimum. However, a more detailed analysis of such calculations shows that the curvature around energy minimum is way off the experimental one, and for large distances between the adsorbate and the substrate LDA calculations do not give the typicall $d^{-6}$ behavior of the tail of binding energy dependence upon distance.  

\begin{figure}[htb]
\begin{center}
\includegraphics[scale=0.4,clip=true]{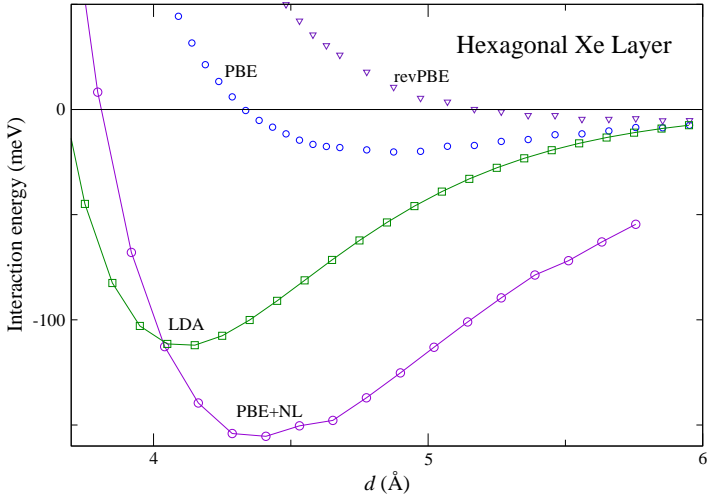}
\end{center}
\caption{Binding energy of a Xenon layer as a function of the layer lattice constant for different energy functionals.} 
\label{xelayerenergy}
\end{figure}

The calculation of the Xenon monolayer is numerically more challenging than the calculation of a Krypton dimer due to the  fact that one has to repeat the charge density in space and calculate nonlocal interaction with even further repeated densities until one reaches the convergence of energy differences.

\section{Conclusions}
We start with a few comments on the state of development of our code. We have approached the problem of calculating and storing the kernel $\phi({\bf r},{\bf r'})$ in Eq.~(\ref{integral}) from the pragmatic standpoint. We have determined the number of points and the limits of the integral, as well as the number of divisions of $D$ and $\delta$ by checking that further refinement does not change the results. We have not tried very hard to optimize the size and organization of the lookup table in order to save computational resources. Also, we are aware of the possibility to use the asymptotic form of the kernel for large $D$ for significantly improving the summation over repeated cells in the case of periodic problems. We expect to address these issues in future work.

We have developed a massively parallel code capable for performing vdW-DF theory calculations on top of largest DFT calculations feasible today. The facts that present standard DFT codes can not cope successfully with nonlocal correlation effects and that our code can be used with densities obtained from almost any existing DFT code makes our {\textit {JuNoLo}} code a valuable tool for the whole \textit{ab initio} community.

%\section{Acknowledgements**}
\section{Acknowledgements}
Calculations were performed on the JUMP and Blue/Gene supercomputers at the Forschungscentrum J\"ulich, Germany.
Two of us, P.~L. and S.~B., thank the Deutsche Forschungsgemeinschaft (DFG)
(Priority Programme ``Molecular Magnetism'') and Alexander von Humboldt foundation. 
N.~A. acknowledges the support of Japan Society for the Promotion of Science.
R.~B. acknowledges the support of MSES of the Republic of Croatia through project No. 098-0352828-2836.

\section{Appendix A. sample files}

Input file for a calculation of Xenon monolayer example is given below.\\
\bf{Input file:} \\
\\
\tt
\noindent out\_Xe\_layer\_3.5-dens3d~~\# charge density file\\
4~~~~~~~~~~~~~~~~~~~~~~~\# periodicity\\
64~~~~~~~~~~~~~~~~~~~~~~\# derivation order\\
0~~~~~~~~~~~~~~~~~~~~~~~\# treating derivations\\
fft\_derivs~~~~~~~~~~~~~~\# derivations file\\
1e-4~~~~~~~~~~~~~~~~~~~~\# trimming value\\
0~~~~~~~~~~~~~~~~~~~~~~~\# calculate kernel 1 or not 0\\
kernel.txt~~~~~~~~~~~~~~\# file with kernel parameters\\
1~~~~~~~~~~~~~~~~~~~~~~~\# perx\\
1~~~~~~~~~~~~~~~~~~~~~~~\# pery\\
0~~~~~~~~~~~~~~~~~~~~~~~\# perz\\
\rm

The output file for the Xenon monolayer example is given below.\\
\bf{Output file:}\\
\\
\tt
\noindent Job started on             2  processors.\\
 using input file input\_Xe\_Layer\_3.5                                          \\
 input density file \\
 out\_Xe\_layer\_3.5-dens3d                                     \\
 periodicity             4\\
 derivation order            64\\
 file used for gradientfft\_derivs                    \\
 trimming limit    1.0000000000000000E-004\\
 perx,pery,perz             1            1            0\\
 calculating gradient by means of fft\\
  bufer\_size\_pp=           100\\
  bufer\_size=           200\\
  nphi0=           100\\
  r\_cutoff=     2000.000000000000       AA\\
 nx,ny,nz        28       28      160\\
       x0,y0,z0         0.00000000        0.00000000        0.00000000\\
       ax,ay,az         7.01525312        4.05025828        0.00000000\\
       bx,by,bz         7.01525312       -4.05025828        0.00000000\\
       cx,cy,cz         0.00000000        0.00000000       49.64206163\\
       dx,dy,dz         0.25054475       -0.14465208        0.31026289\\
     b1x,b1y,b1z       -0.44782314       -0.77565242        0.00000000\\
     b2x,b2y,b2z       -0.44782314        0.77565242        0.00000000\\
     b3x,b3y,b3z        0.00000000        0.00000000       -0.12656979\\
             dV         0.02248898\\
 total number of points        125440\\
 number of points per processor         62720\\
 starting=================\\
 y 2008 d  3 m  9\\
12: 8:57:224\\
 running on             2  CPUs\\
 master is calculating gradient using fft\\
 to calculate fft*******************\\
elapsed\_time  0 D:  0 H:  0 M:  1 S:  907 MS\\
 TOTAL NUMBER OF ELECTRONS IS     29.95081769527724     \\
 min and max density on master   -1.6042787840499999E-006 \\
    15.41315504760000     \\
 to distribute density*******************\\
elapsed\_time  0 D:  0 H:  0 M:  0 S:  277 MS\\
 to calculate local energy*******************\\
elapsed\_time  0 D:  0 H:  0 M:  0 S:  104 MS\\
 total number of points after trimming         24617\\
 NUMBER OF ELECTRONS AFTER TRIMMING     29.94276421010878     \\
 q0min,q0max   0.2468401900000000         8.096420200000001     \\
 to load and distribute trimmed points*******************\\
elapsed\_time  0 D:  0 H:  0 M:  1 S:  274 MS\\
 to load kernel*******************\\
elapsed\_time  0 D:  0 H:  0 M:  0 S:  163 MS\\
 nphi,qmin\_cut,phi0\_analitic          100    1.100000000000000      \\
    2.500000000000000     \\
 px,py,pz            3            3            3\\
 to calculate phi0*******************\\
elapsed\_time  0 D:  0 H:  2 M: 17 S:   19 MS\\
 to calculate 0th unit cell energy*******************\\
elapsed\_time  0 D:  0 H:  0 M: 50 S:  517 MS\\
 to calculate energy*******************\\
elapsed\_time  0 D:  0 H:  9 M: 46 S:  523 MS\\
 ------ENERGIES  (eV) ----------------\\
 Ex\_lda Ec\_lda Exc\_lda \\
       TOTAL\_LDA     -783.10552037      -59.77262220     -842.87814257\\
 Ex\_pbe Ec\_pbe Exc\_pbe \\
       TOTAL\_PBE     -810.50311278      -42.20550892     -852.70862170\\
 X\_rev\_PBE   -811.6072505786024     \\
 E\_OFFDIAG\_nlc E\_DIAG\_nlc E\_TOTAL\_xc\_nlc\\
        TOTAL\_NL       -3.47092761        9.63845229        6.16752468\\
 needed number -Ec\_pbe+Ec\_lda+Enl=   -11.39958860142493     \\
 ------END----------------------------\\
 to collect points*******************\\
elapsed\_time  0 D:  0 H:  0 M:  2 S:  102 MS\\
 GAME OVER=================\\
 y 2008 d  3 m  9\\
12:21:57: 71\\
\rm

%\section{References}

%** These sections are optional

\end{document}